\documentclass[aps,pre,twocolumn, superscriptaddress]{revtex4}

\usepackage{graphicx}
\usepackage{amssymb}
\usepackage{amsmath}
\usepackage{color}

\begin{document}
\title{Entropy and Kinetics of Point-Defects in Two-Dimensional Dipolar Crystals}

\author{Wolfgang Lechner}
\email{w.lechner@uibk.ac.at}
\affiliation{Institute for Quantum Optics and Quantum Information, Austrian Academy of
Sciences, 6020 Innsbruck, Austria}
\affiliation{Institute for Theoretical Physics, University of Innsbruck, 6020 Innsbruck,
Austria}

\author{David Polster}
\affiliation{Department of Physics, University of Konstanz, D-78457 Konstanz, Germany}

\author{Georg Maret}
\affiliation{Department of Physics, University of Konstanz, D-78457 Konstanz, Germany}

\author{Christoph Dellago}
\affiliation{Faculty of Physics, University of Vienna, Boltzmanngasse 5, 1090 Vienna, Austria}

\author{Peter Keim}
\affiliation{Department of Physics, University of Konstanz, D-78457 Konstanz, Germany}

\date{\today}

\begin{abstract}
We study in experiment and with computer simulation the free energy and the kinetics of vacancy and interstitial defects in two-dimensional dipolar crystals. The defects appear in different local topologies which we characterize by their point group symmetry; $C_n$ is the n-fold cyclic group and $D_n$ is the dihedral group, including reflections. The frequency of different local topologies is not determined by their almost degenerate energies but dominated by entropy for symmetric configurations. The kinetics of the defects is fully reproduced by a master equation in a multi-state Markov model. In this model, the system is described by the state of the defect and the time evolution is given by transitions occurring with particular rates. These transition rate constants are extracted from experiments and simulations using an optimisation procedure. The good agreement between experiment, simulation and master equation thus provides evidence for the accuracy of the model.
\end{abstract}
\pacs{}

\maketitle

\section{Introduction}

The microscopic dynamics and interaction of defects, like dislocations, vacancies and interstitials are key to a variety of macroscopic phenomena of materials \cite{TAYLOR,NELSON}. In two dimensional systems \cite{Kusner1994,Marcus1996,Zahn2000,Han2008,Deutschlaender2014} the peculiar two step melting is a result of dislocation  and disclination interactions and melting is mediated by the formation and subsequent dissociation of dislocation pairs \cite{KTHNY,Frenkel1979,Strandburg1983,Jaster1998,Sengupta2000,Mak2006,Shiba2009,Bernard2011}. Dislocations pairs in 2d-crystals also form as the result of spontaneous clustering of interstitals and vacancies introduced into the systems \cite{LECHNERDEFECTSTRINGS}. The dynamics and interaction of interstitials and vacancies is even related to exotic phases such as supersolidity \cite{FATE,DEFECTINDUCED}. Nevertheless, even though individual defects and their displacement fields are well described by elasticity theory, the precise kinetics and the sign of the interaction cannot be captured by the theory \cite{LDSoftMatter1,LDSoftMatter2}. The open question is how the defect kinetics emerges from the non-linear effects near the defect centers. Video microscopy of two dimensional colloidal crystals together with optical tweezers now allow to investigate such nonlinear effects with single particle resolution in real time \cite{CONFOCAL,PERTSINIDIS_NJP,PERTSINIDIS_PRL,MARET,MELTINGREVIEW}.

Here, we show that the dynamics of defects in two-dimensional dipolar crystals can be fully described by a sequence of jumps between states which are defined by the local displacements in the vicinity of the defect centers. The equilibrium probabilities of the states (i.e. the populations) are a result of the interplay between entropic and energetic contributions. We find that the different contributions can be understood quantitatively from statistical mechanics by a harmonic expansion of the energy around the minima corresponding to the various defect states. The kinetics of the defects follows a master equation for which we measure the transition matrix from a long experimental trajectory with the aid of an optimization routine. The results from experiments are compared with the results from Monte Carlo simulations. The equilibrium probabilities from simulation and experiment are in qualitative good agreement but show systematic differences for stiff crystals.

\begin{figure}[ht]
\centerline{\includegraphics[width=9cm]{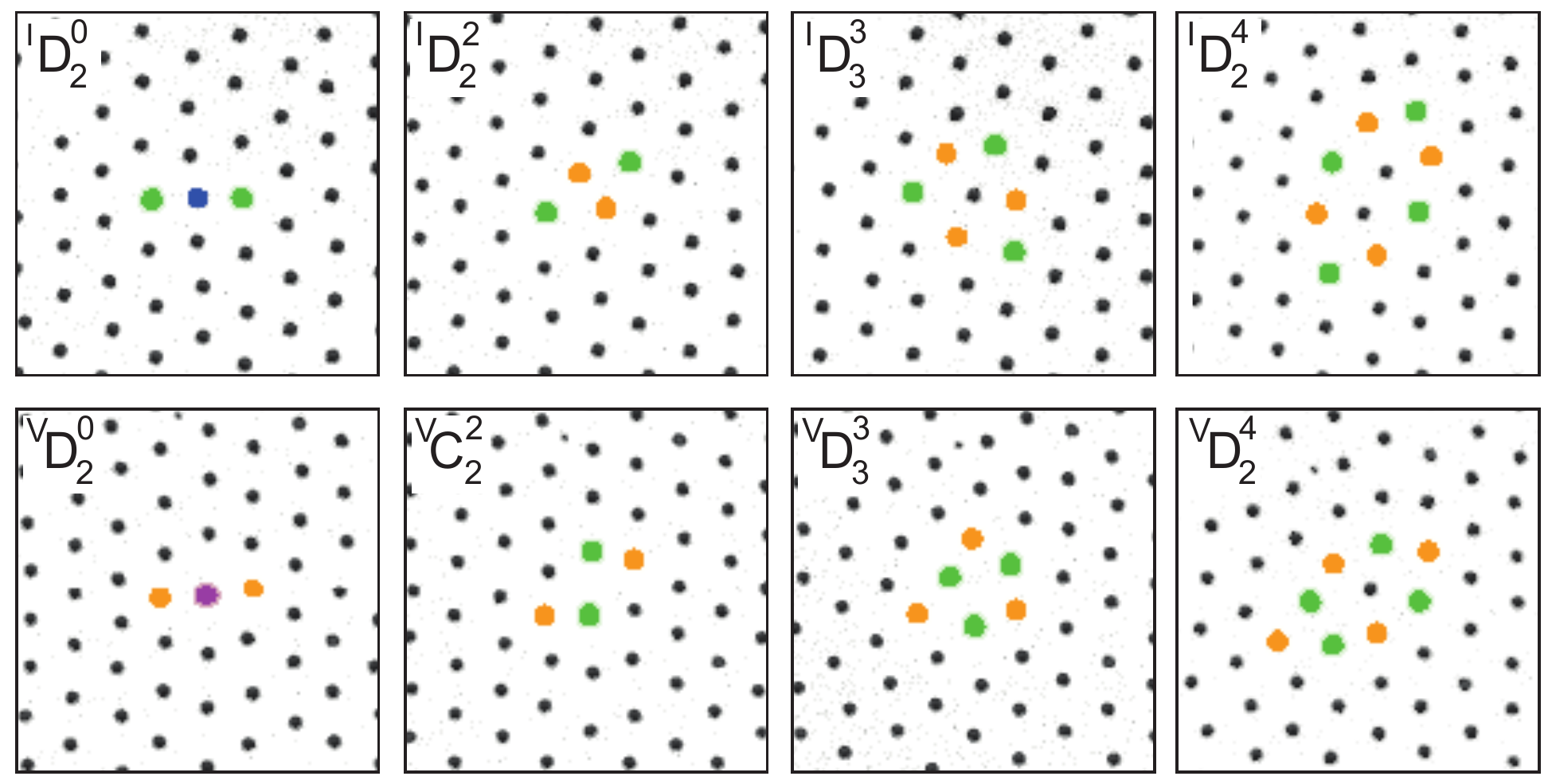}}
\caption{Typical snapshots of the colloidal crystal obtained from the experiment containing an interstitial (upper row) and a vacancy (lower row), respectively. The motion of the defects consists of a sequence of transitions between states with different symmetries given by the dihedral $D_n$ and cyclic $C_n$ point groups in 2D. The upper left index denotes interstitial versus vacancies and the upper right counts the numbers of seven-fold particles (in case of interstitials) and five-fold particles (in case of vacancies). Particles with $4$, $5$, $7$, and $8$ neighbors are shown in blue, orange, green, and purple, respectively, and particles with $6$ neighbors are shown in black. Note that the symmetry of the most frequent low temperature configurations (second column) differ for interstitials and vacancies. The vacancies show only cyclic symmetry for low temperatures.}
\label{fig:states}
\end{figure}

\section{Experimental Setup and Model}

We consider a system of super-paramagnetic colloidal particles ($4.5 \mu$m in diameter) confined by gravity to two dimensions on a flat interface. Two situations were realized, a flat water/air interface in hanging droplet geometry and a solid interface of a glass substrate. The only difference in both realizations is a slightly enhanced self diffusion coefficient of colloidal particles at the water/air interface on the expense of an increased equilibration time after mounting the sample compared to the solid substrate. The latter is due to a nontrivial regulation of the curvature of the droplet \cite{Ebert09}. Video microscopy and digital image analysis provides the position of the particles at a frame rate of about $1$ Hz which is fast compared to the Brownian timescale of $50$ sec. The colloidal ensemble is described by the Hamiltonian
\begin{equation}
H = \sum_i \frac{p_i^2}{2m} + \sum_{i<j} V(r_{ij}),
\label{equ:hamiltonian}
\end{equation}
with the pair interaction for particles at distance $r$
\begin{equation}
\beta V(r) = \frac{3^{3/4}\Gamma}{(2\pi)^{3/2}} \left( \frac{a}{r} \right)^3.
\label{equ:interaction}
\end{equation}
Here, $\beta = 1/k_B T$ with the Boltzmann constant $k_B$ and temperature $T$. The distance is given in units of $a$, the average inter-particle distance of the triangular lattice. The dimension-less parameter  $\Gamma = \beta (\mu_0/4\pi)(\chi H)^2 (\pi \rho)^{3/2}$ defines the phase behavior of the system and can be interpreted as an inverse temperature, tuned with the magnetic field $H$. Here, $\mu_0$ is the permeability of vacuum, $\chi$ is the magnetic susceptibility of the particles and $\rho$ is the 2d particle density. The experimental setup is described in detail in Refs. \cite{Ebert09,MARET,LECHNERDEFECTSTRINGS}. On the solid substrate a vacancy is prepared by trapping a colloid with an optical tweezer (20mW, 100x tweezer-objective from above the sample, NA 0.73, $\textrm{Ar}^{+}$-Laser) and pulling it alongside a lattice line out of the field of view (at least more than 15 lattice spacings). Correspondingly, an interstitial is created by pulling a particle from the far field into the center of the field of view of an otherwise defect free crystal. Samples confined at the water/air interface offers additionally the possibility to shoot particles out of the interface by light pressure with a strong laser pulse (500mW). Since the sound velocity is of the order of a few mm/s the distorted lattice relaxes rapidly and measurements are started after about a minute. Positional data were taken at three different interactions strength $\Gamma = 120,140,156$ with more than 3,000 configurations in the case of $\Gamma = 120$ and $156$ and more than 20,000 configurations in the case of $\Gamma = 140$ for both interstitials and vacancies. For $\Gamma = 140$ the creation of the defect was repeated frequently, annealing the crystal and equilibrating the systems in between for several days up to a few weeks.

\section{Defect Classification and Patterns}

In equilibrium, the vacancies and interstitials exist in different, almost degenerate states with different topologies \cite{Jain00,GLIDING}. The most frequently appearing vacancy and interstitial states are shown in Fig. \ref{fig:states} and we classify the defects according to their point groups in two dimension. $C_n$ denotes the cyclic group with n-fold rotational symmetry whereas $D_n$ is the dihedral group with additionally $n$ mirror axes. The upper left index denotes vacancies or interstitials and the upper right index counts the number of dislocations involved in the defect. In 2D a dislocation is a pair of a five- (colored orange in Fig.\;\ref{fig:states}) and seven-fold (colored green) coordinated particles characterized by a burgers-vector. The neighbor numbers in the vicinity of the defect centers are determined using a Voronoi construction \cite{VORONOI}. The dissociation of dislocations is known to drive the melting transition in 2D \cite{KTHNY}. The most frequent defect configurations are listed in table \ref{tbl:definition} and the index $i$ is introduced to label the defects in the  formula below and for the computations of  transition rates. Configurations with larger numbers of dislocations exist and are numbered with \textit{others} ($i = 0$). Note, that the number of such defects vanishes for large $\Gamma$ (see dashed lines in Fig. \ref{fig:probabilities}).

\begin{table}
\begin{tabular}{c|c|c|c|c|c}
Defect & \;i\; & \#4 & \#5 & \#7 & \#8 \\
\hline
$^ID\,_2^0$ & 1 & 1 & 0 & 2 & 0 \\
$^ID\,_2^2$ & 2 & 0 & 2 & 2 & 0 \\
$^ID\,_3^3$ & 3 & 0 & 3 & 3 & 0 \\
$^VD\,_2^4$ & 4 & 0 & 4 & 4 & 0 \\
\hline
$^VD\,_2^0$ & 1 & 0 & 2 & 0 & 1 \\
$^VC\,_2^2\,/\,^VD_2^2$ & 2 & 0 & 2 & 2 & 0 \\
$^VD\,_3^3$ & 3 & 0 & 3 & 3 & 0 \\
$^VD\,_2^4$ & 4 & 0 & 4 & 4 & 0 \\
\end{tabular}
\caption{Classification of interstitial (I) and vacancy (V) states based on the number of particles with 4, 5, 7 and 8 neighbors in the vicinity of the defect.}
\label{tbl:definition}
\end{table}

\begin{figure}[htb]
\centerline{\includegraphics[width=9cm]{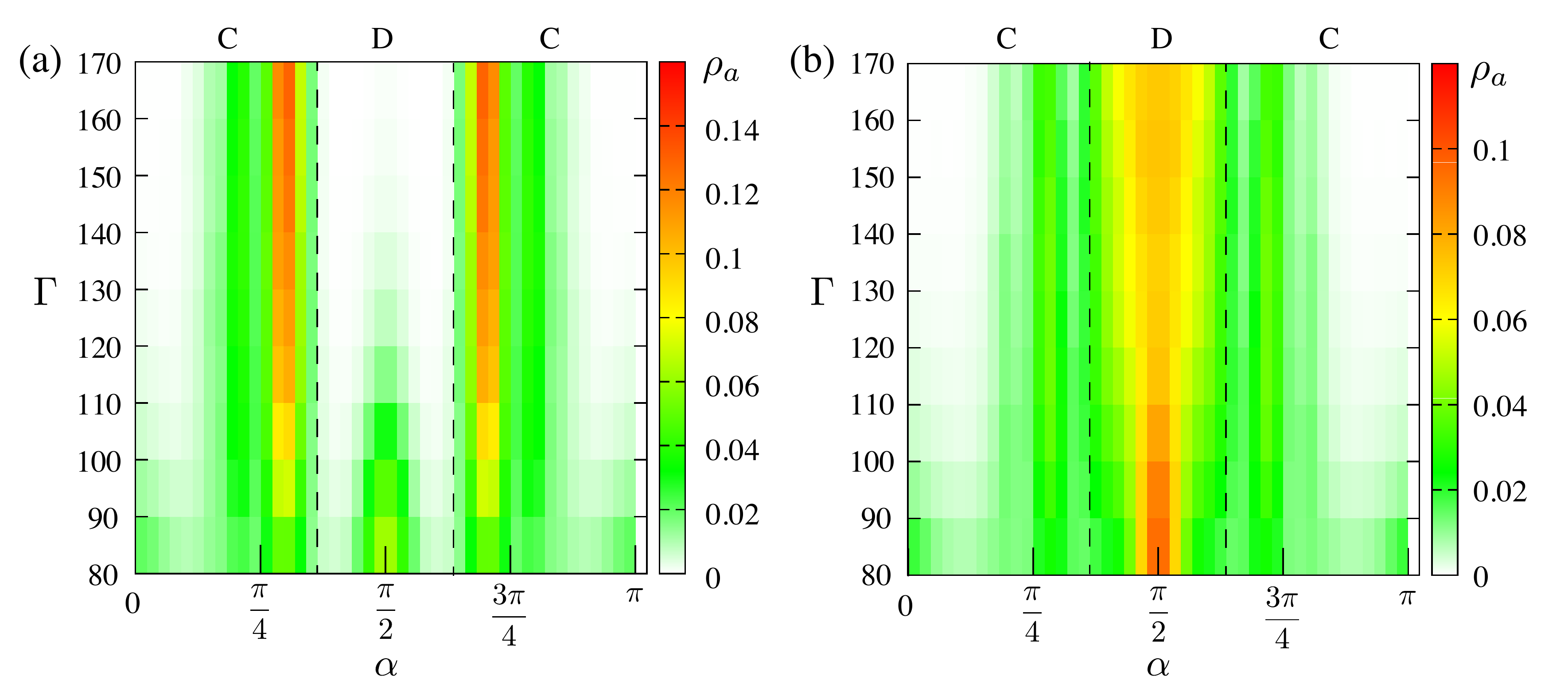}}
\caption{Probability density of the angle $\alpha$ between the 5-5 and 7-7-bonds for vacancies (left) and interstitials (right) consisting of two dislocations for different values of $\Gamma$. Red indicates high probabilities. For $\alpha = \pi/2$ the symmetry is dihedral $D_2$. The vacancies are peaked at $\alpha \approx 90^\circ \pm 38^\circ$ for low temperatures (large $\Gamma$) with cyclic symmetry $C_2$.}
\label{fig:bondangles}
\end{figure}

\begin{figure}[htb]
\centerline{\includegraphics[width=8cm]{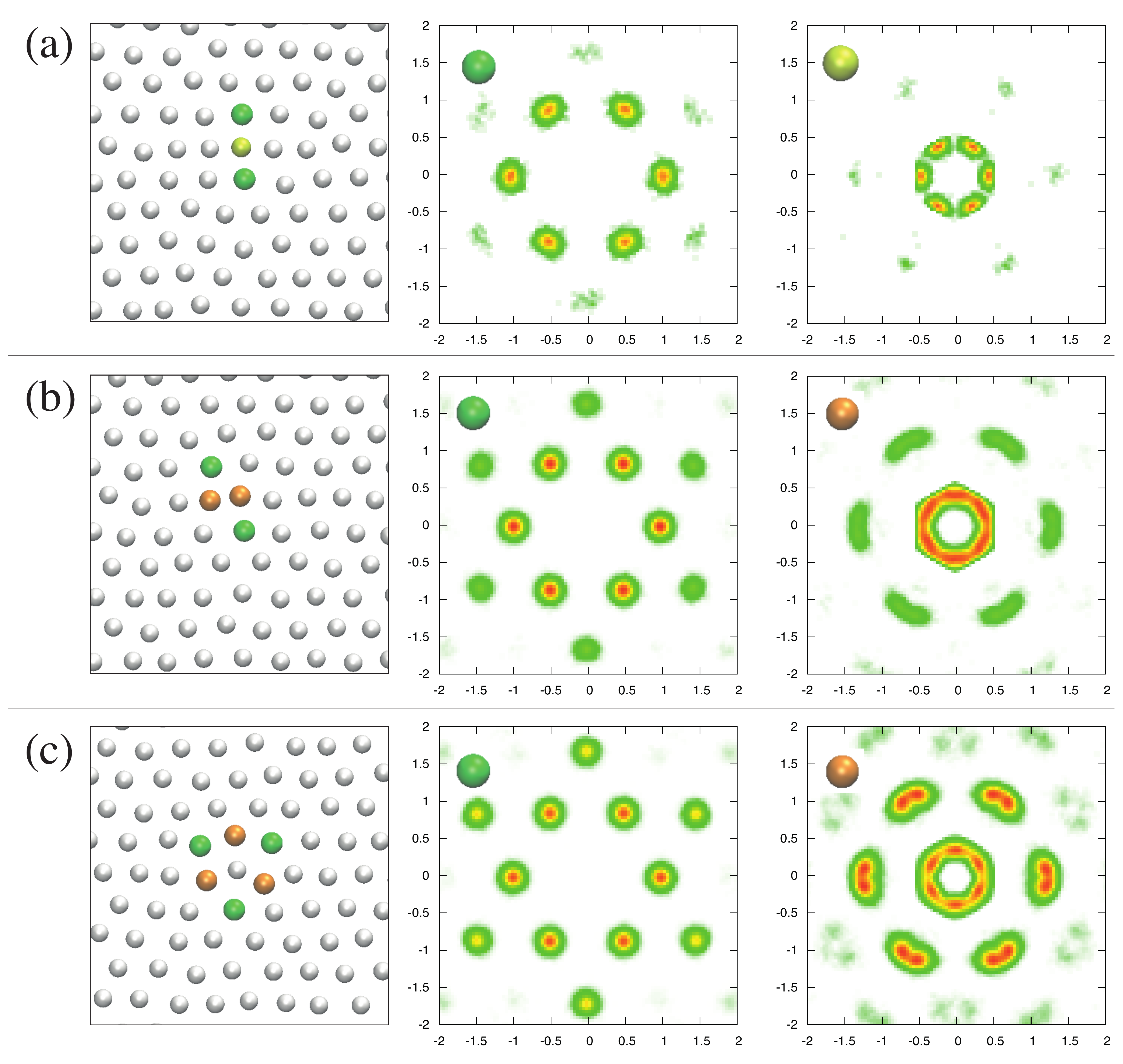}}
\caption{Relative positions of $4$, $5$, and $7$-coordinated particles in the vicinity of an interstitial defect with respect to the defect position on the underlying triangular lattice from computer simulations. The configurations (left  column)  $^ID_2^0$ (a), $^ID_2^2$ (b) and $^ID_3^3$ (c) consist of at least two 5-coordinated particles (green). The $^ID_2^0$ state consists of an additional 4-coordinated particle (yellow) while $^ID_2^2$ and $^ID_3^3$ state consist of additional 7-coordinated particles (orange). The relative positions of 5-coordinated (middle column) and 4- and 7-coordinated particles (right column) are measured with respect to the closest lattice point of the defect center and depicted as a heat map. Red indicates large probability and white low probabilities. Clearly, the 5-coordinated particles occupy always the same 12 lattice points with different weights depending on the state. The 5 and 7-coordinated particles also follow different patterns. }
\label{fig:interstitials_lattice}
\end{figure}
\begin{figure}[htb]
\centerline{\includegraphics[width=8cm]{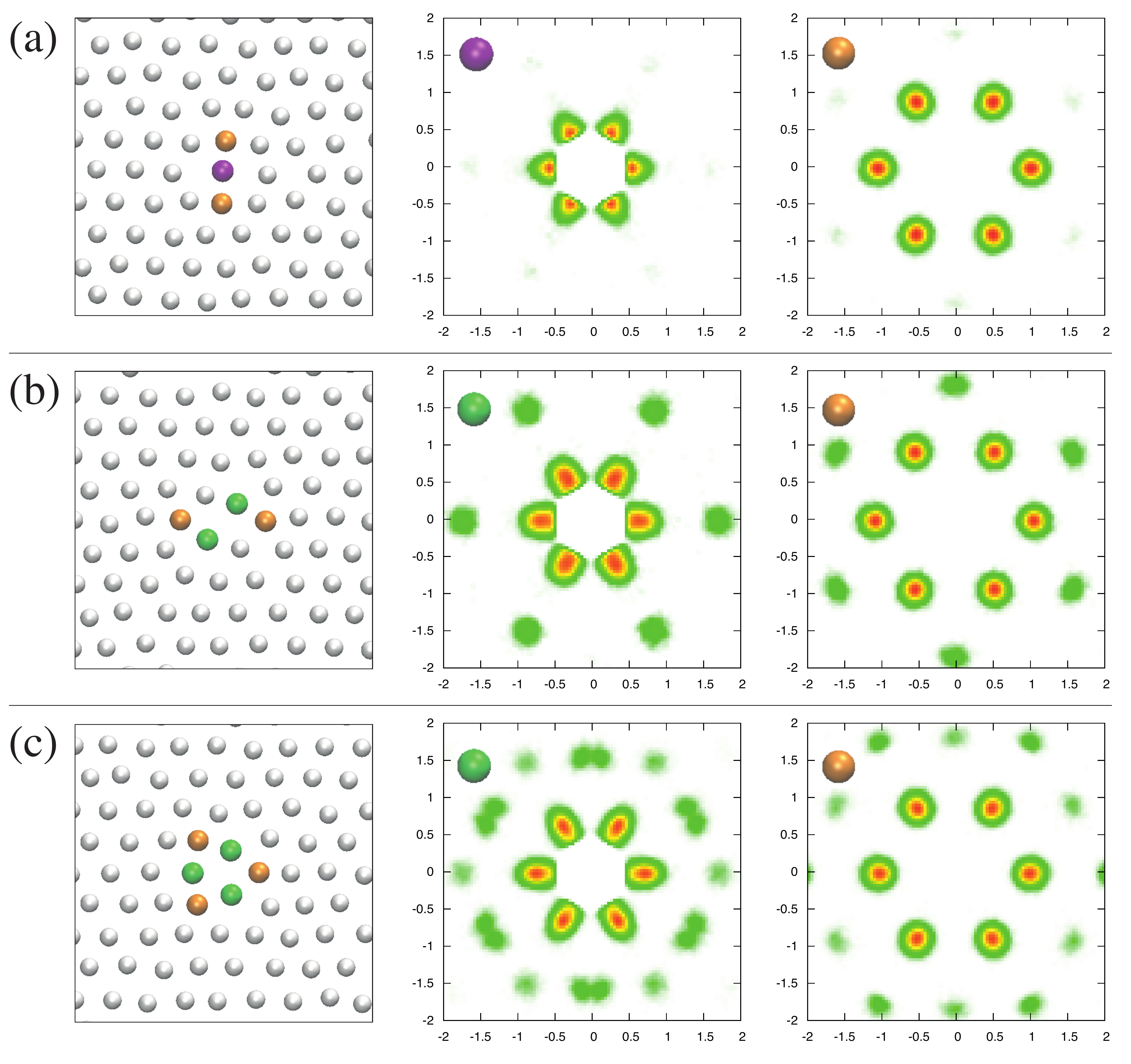}}
\caption{Relative positions of $5$, $7$, and $8$-coordinated particles in the vicinity of a vacancy defect with respect to the defect position on the underlying triangular lattice from computer simulations as described in the caption of Fig. \ref{fig:interstitials_lattice}. A vacancy in state $^VD_2^0$ (a),  $^VC_2^2$ (b), or $^VD_3^3$ (c) consists of at least two 7-coordinated particles (orange) and either a single 8-coordinated (violet) particle (a) or additional 5-coordinated particles (green) (b) and (c).  In all states, the 7-coordinated particles are found at the same 12 positions (right) while the 8 and 5-coordinated particles are found at various other positions. }
\label{fig:vacancies_lattice}
\end{figure}
Interestingly, the typical symmetry of the configuration consisting of two dislocations (second column in Fig.\;\ref{fig:states}) is not the same for interstitials and vacancies. To quantify the relative probability of dihedral and cyclic configurations we introduce the following procedure. We connect the five-fold particles with a line and the seven-fold ones too and measure the angle $\alpha$ between the two axis. For $\alpha = \pi/2 \pm \pi/8$ the pattern is classified as rhombic ($D_2$ symmetry) and otherwise classified as cyclic ($C_2$ symmetry).  Fig.\;\ref{fig:bondangles} shows the probability density as heat map for vacancies (a) and interstitials (b). Red corresponds to large probabilities and green to low ones. For $\Gamma \geq 120$ the distribution is sharply peaked at about $\alpha \approx \pi/2$ for the vacancies and the probability of the $D_2$ symmetry tends to zero. The ratio for $D_2$ symmetry increases to $30 \%$ at $\Gamma = 80$. Having this in mind we keep the notation $^VC_2^2$ for vacancies consisting of two dislocations in the following plots for clarity. For interstitials Fig.\;\ref{fig:bondangles}(b) the distribution is peaked at $\alpha \approx \pi/2$ corresponding to dihedral symmetry $D_2$ for all interaction strength. Nonetheless the distributions widens for lower temperatures which might indicate a separate configuration $C_2$. Using the same cutoff as for vacancies the ratio varies form $50\% : 50\%$ at $\Gamma = 80$ to $75\%$ $^ID_2$ and $25\%$ $^IC_2$ at $\Gamma = 170$.

To visualize the configurations with respect to the underlying lattice of the crystal, Figures \ref{fig:interstitials_lattice} and \ref{fig:vacancies_lattice} show the probabilities of positions of particles with respect to the defect centers. All patterns reflect the symmetry of the crystal but the relative positions of particles with 4, 5, and 7 neighbors for the interstitials are arranged in particular patterns shown in Fig. \ref{fig:interstitials_lattice} and the 5, 6, and 8 neighbored particles of the vacancies are shown in Fig. \ref{fig:vacancies_lattice}. Remarkably, the positions of the 5-coordinated particles in an interstitial defect and the positions of 7-coordinated particles are all degenerate within the 12 points in a star-like pattern.

\section{Equilibrium Defect Populations}

\begin{figure}[b]
\centerline{\includegraphics[width=7cm]{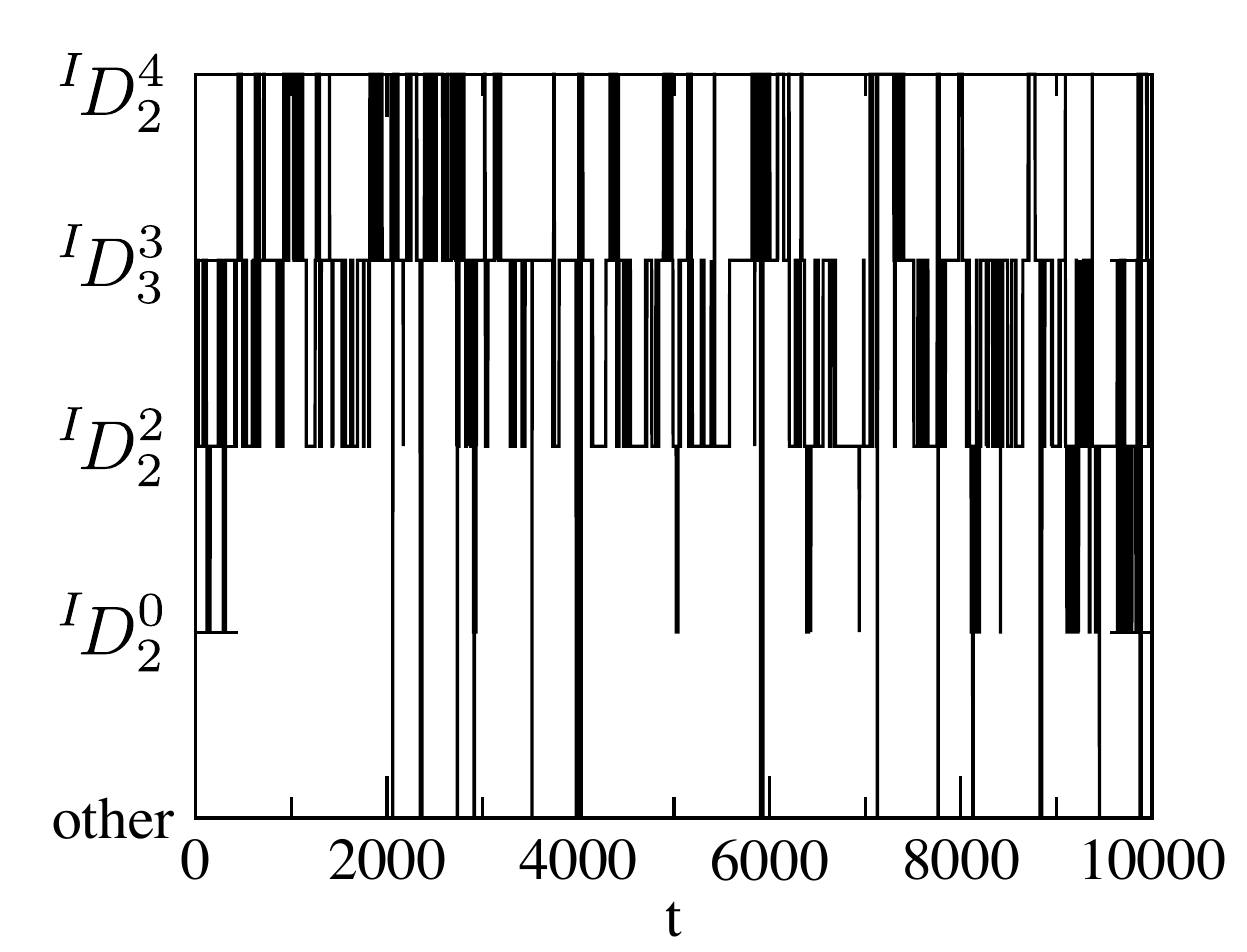}}
\caption{Typical trajectory of the state as a function of time (in units of Monte Carlo time steps) of an interstitial at $\Gamma=160$. The state is identified in each time step from the neighbor numbers of particles in the vicinity of the defect according to Table I.}
\label{fig:trajectory}
\end{figure}

\begin{figure}[htb]
\centerline{\includegraphics[width=9cm]{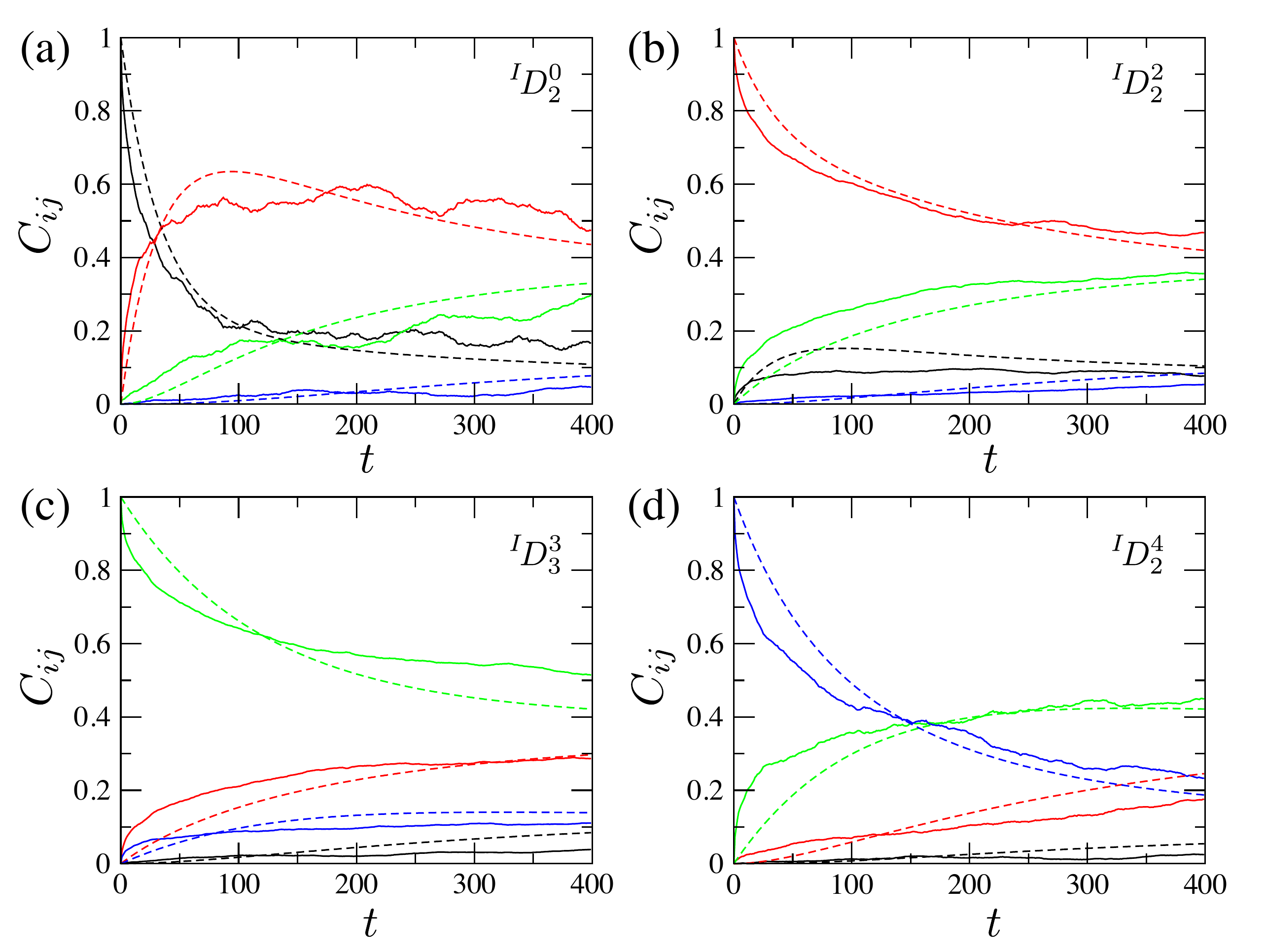}}
\caption{Time correlation functions $C_{ij}(t)$ for interstitials obtained from experiment (solid lines) in comparison to the result of the optimization procedure of $K_{ij}$ (dashed) for $\Gamma=140$. The symbols in the top right corner of each panel indicate the initial state $j$  and colors indicate the final state $i$, where black corresponds to $^ID_2^0$, red to $^ID_2^2$, green to $^ID_3^3$ and blue to $^VI_2^4$. For instance, the black lines in the top left panel represent the probability to find the defect in the $^ID_2^0$ configuration at time $t$ given that it was in $^ID_2^0$ at time $t=0$, while the red lines represent the probability to find it in $^ID_2^2$ at time $t$.}
\label{fig:ciji_exp}
\end{figure}

\begin{figure}[htb]
\centerline{\includegraphics[width=9cm]{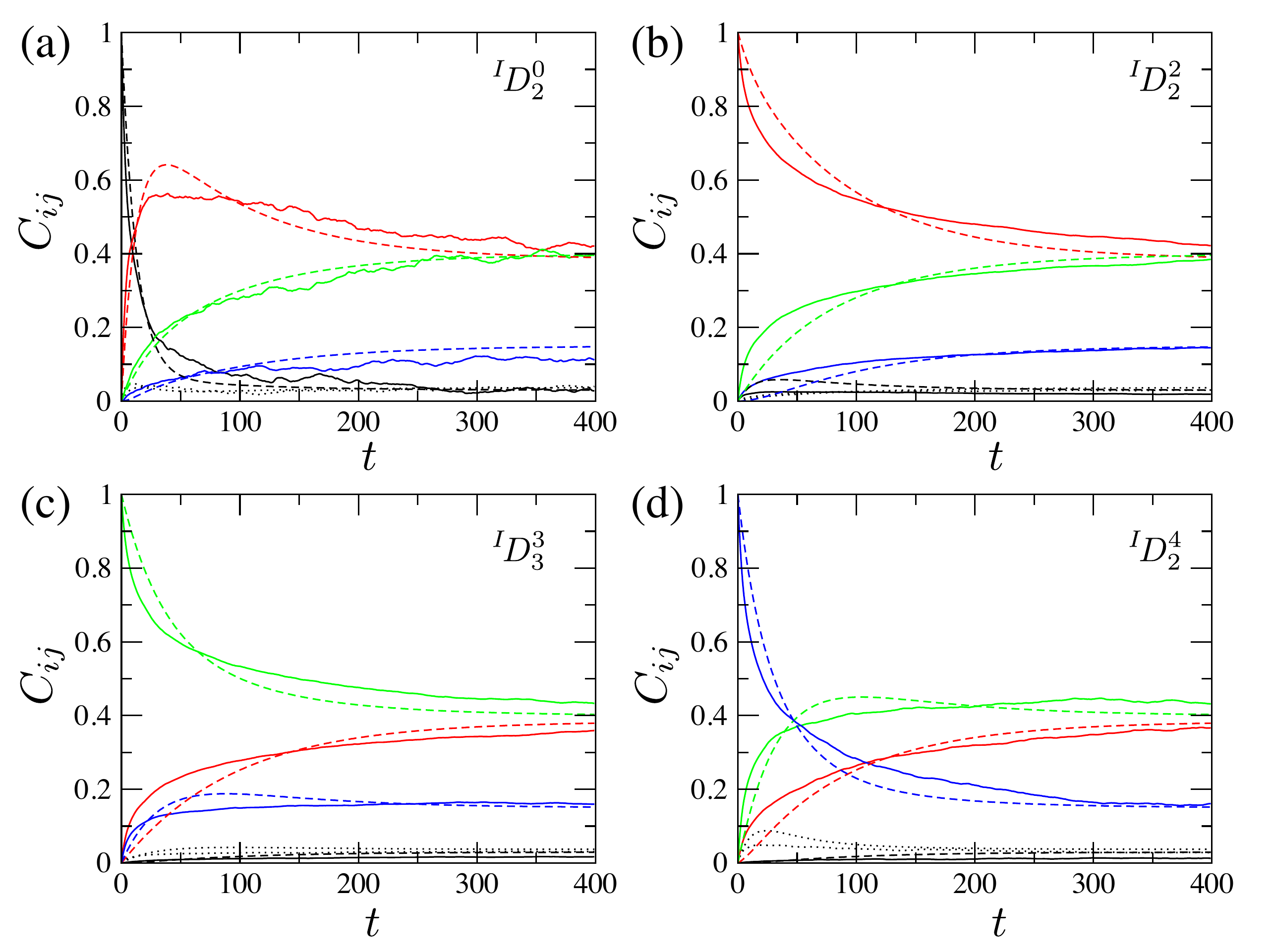}}
\caption{Correlation functions $C_{ij}(t)$ for interstitials obtained from simulation (solid lines) and from the optimization procedure of $K_{ij}$  (dashed). Parameters and colors as in Fig. \ref{fig:ciji_exp}.  }
\label{fig:ciji}
\end{figure}

\begin{figure}[htb]
\centerline{\includegraphics[width=9cm]{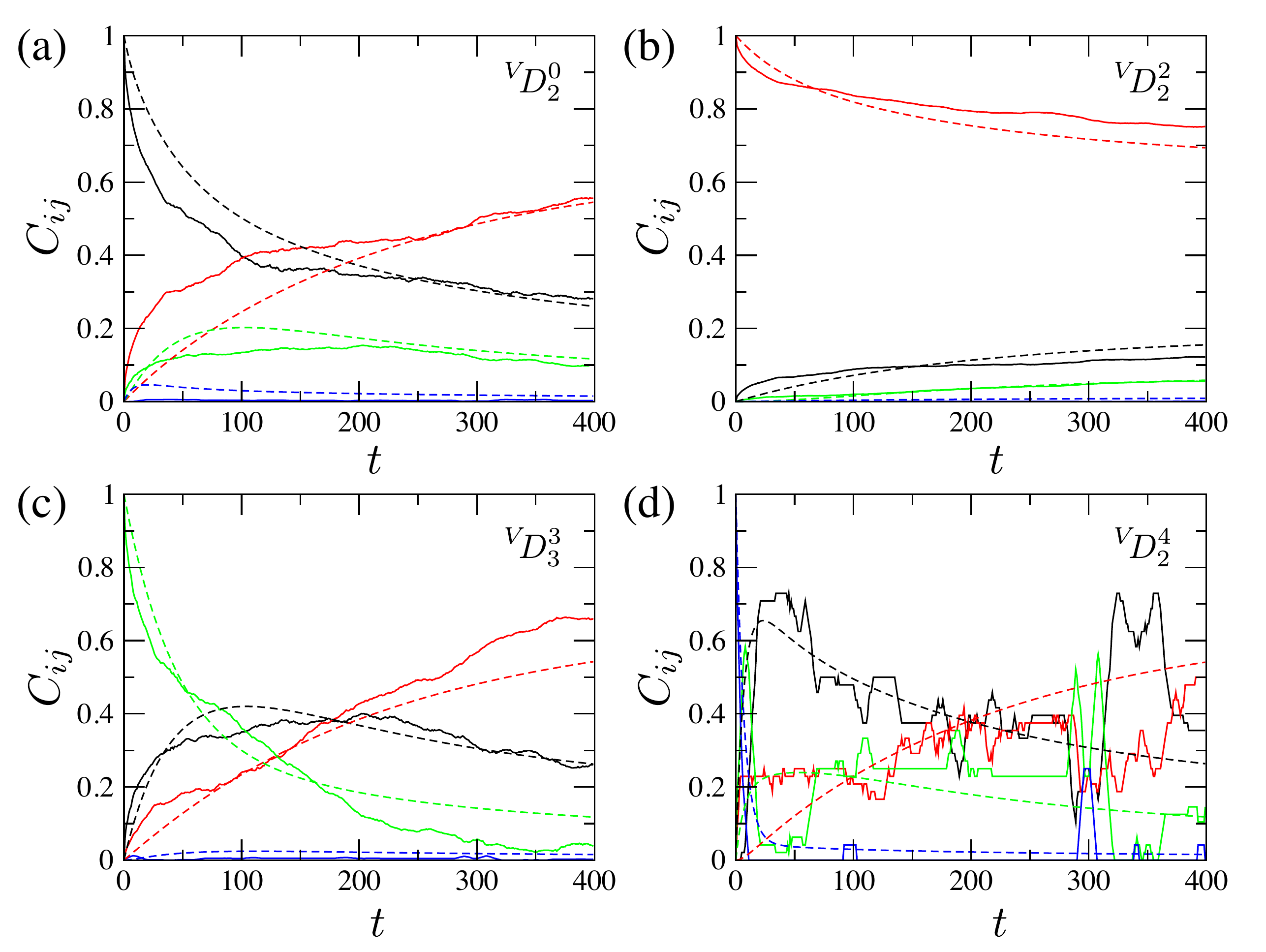}}
\caption{Correlation functions $C_{ij}(t)$ for vacancies obtained from  experiment (solid line) and from the optimization procedure of $K_{ij}$  (dashed). Parameters and colors as in Fig. \ref{fig:ciji_exp}}
\label{fig:exp_cijv}
\end{figure}

\begin{figure}[htb]
\centerline{\includegraphics[width=9cm]{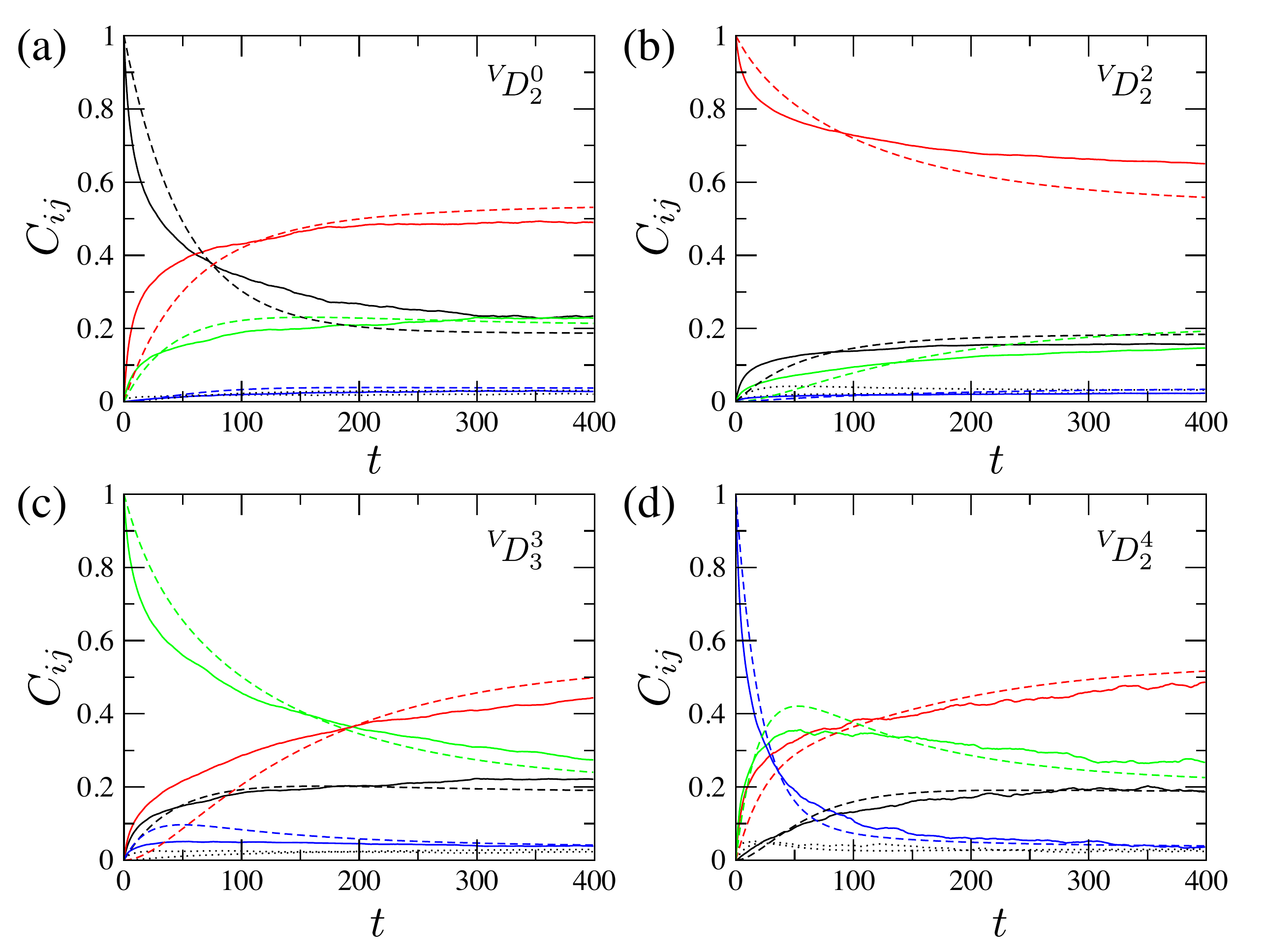}}
\caption{Correlation functions $C_{ij}(t)$ for vacancies obtained from simulation (solid line) and from the optimization procedure of $K_{ij}$  (dashed). Parameters and colors as in Fig. \ref{fig:ciji_exp} }
\label{fig:cijv}
\end{figure}

Following a trajectory $\mathbf{x}(t)$ of the system in experiment one can use the classification above to identify a trajectory of states. Here, $\mathbf{x}(t)$ denotes the configuration of the system, including the positions of all particles, at time $t$. In computer simulations, $\mathbf{x}(t)$ is the sequence of Monte Carlo configurations. For illustration, a typical trajectory taken from computer simulations is depicted in Fig. \ref{fig:trajectory}.
From the trajectory, we determine the probability $P_i$ of finding the defect in state $i$ with
\begin{equation}
P_i = \langle h_{i}[\mathbf{x}(t)] \rangle = \lim\limits_{\tau \rightarrow \infty}{\frac{1}{\tau} \int_0^\tau dt' h_i[\mathbf{x}(t')] }.
\label{equ:prob}
\end{equation}
The indicator function $h_i(\mathbf{x})$ is defined as
\begin{equation}
h_i(\mathbf{x})=
\begin{cases}
    1 & \text{if {\bf x} is in state $i$}, \\
    0              & \text{otherwise}.
\end{cases}
\label{equ:indicator}
\end{equation}
The probabilities are normalized to $\sum_i P_i = 1$ for vacancies and interstitials separately. The population $P_i$ equals the fraction of time the defects spends in state $i$ in equilibrium. This probability is estimated from a finite number of configurations,
\begin{equation}
P_m = \bar{h} = \frac{1}{N} \sum_{i=1}^N h_m[x(i \Delta t)].
\end{equation}
The configurations are sampled in experiment and simulation at constant time intervals. These configurations may therefore be correlated and the error from this estimation, including correlated events is

\begin{equation}
\sigma^2 = \langle (\bar{h} - \langle h \rangle )^2 \rangle = \frac{1}{N^2} \sum_{ij} \langle \delta h(i) \delta h(j)\rangle,
\label{equ:sigma2sum}
\end{equation}
where $\langle h \rangle$ is the ensemble average of $h$, and ${\delta h(i) = h[x(i \Delta t)] - \langle h \rangle}$. The correlation function $\langle \delta h(i) \delta h(j)\rangle$ can be rewritten as

\begin{eqnarray}
H (|i-j|) & = & \langle \delta h(i) \delta h(j)\rangle = \langle h(i) h(j) \rangle - \langle h \rangle^2 = \label{equ:Hij} \\  \nonumber
& = & C_{m,m}(\Delta t_{i-j}) - \langle h \rangle^2.
\end{eqnarray}
Here, $C_{m,m}$ are the auto-correlation functions as depicted in Figs. \ref{fig:ciji_exp} - \ref{fig:cijv}. Note, that in equilibrium $H(i-j) = H(|i-j|)$ and $H(0) = \langle h \rangle (1 - \langle h \rangle)$.  Inserting Eq. (\ref{equ:Hij}) into  Eq. (\ref{equ:sigma2sum}) with $t = |i-j|$ we find

\begin{equation}
\sigma^2 = \frac{1}{N} \left( 2 \sum_{t=0}^{N-1} H(t) - H(0) \right) - \frac{2}{N^2} \sum_{t=1}^{N-1} t H(t).
\end{equation}
In the limit of large $N$, the last term, which scales quadratically and the constant term $H(0)$ can be neglected and the error can be estimated from

\begin{equation}
\sigma^2 \approx \frac{2}{N} H(0) \sum_{t=0}^{N-1}  \frac{H(t)}{H(0)} = \frac{\langle h \rangle (1 - \langle h \rangle)}{N}  \sum_{t=0}^{N-1}  \frac{H(t)}{H(0)}.
\end{equation}
Note that the term $\frac{\langle h \rangle (1 - \langle h \rangle)}{N} = \sigma_0^2$ is the error from $N$ uncorrelated measurements. In a general trajectory correlations increase the error by $2 \sum_{t=0}^{N-1}  \frac{H(t)}{H(0)}$. This factor can be associated with the correlation time $t_c$ in the system by
\begin{equation}
\sum_{t=0}^{N-1}  \frac{H(t)}{H(0)} \approx \frac{1}{\Delta t} \frac{1}{H(0)} \int_0^\infty H(t) dt = \frac{t_c}{\Delta t}.
\end{equation}
Here, the correlation time is $t_c = \frac{1}{H(0)} \int_0^\infty H(t) dt$ and $\Delta t$ is the sampling interval. Combining all this, the error from correlated trajectories can be written as the uncorrelated error multiplied by the correlation time
\begin{equation}
\sigma^2 = \sigma^2_0 \frac{2 t_c}{\Delta t}.
\end{equation}
In the experiments, the trajectory is sampled every $\Delta t = 0.92$ seconds.
The number of measurements differs considerably for the various $\Gamma$ for vacancy and interstitials. In particular, for interstitials $N_{\Gamma=120}^I = 670$ and $N_{\Gamma=140}^I = 31,500$ and for vacancies $N_{\Gamma=120}^V = 4,950$, $N_{\Gamma=140}^V = 23,700$ and $N_{\Gamma=156}^V = 9,900$. The correlation time for different species ranges from $t_c=20$ to $t_c=40$.

In the computer simulations, the trajectory is a sequence taken from Monte Carlo updates. The number of particles is $N_p=26\times30+1 = 781$ for interstitials and $N_p=779$ for vacancies. Interactions are cut off at a radius $r_c = 8 a_0$, where $a_0$ is the lattice spacing. The displacement in each Monte Carlo step are chosen such that the average acceptance rate is approximately $0.5$. Sequences are sampled every $800,000$ Monte Carlo steps which corresponds approximately $1s$ in real time. For all parameters, the total number of measurements is $N=750,000$.

\begin{figure}[htb]
\centerline{\includegraphics[width=7cm]{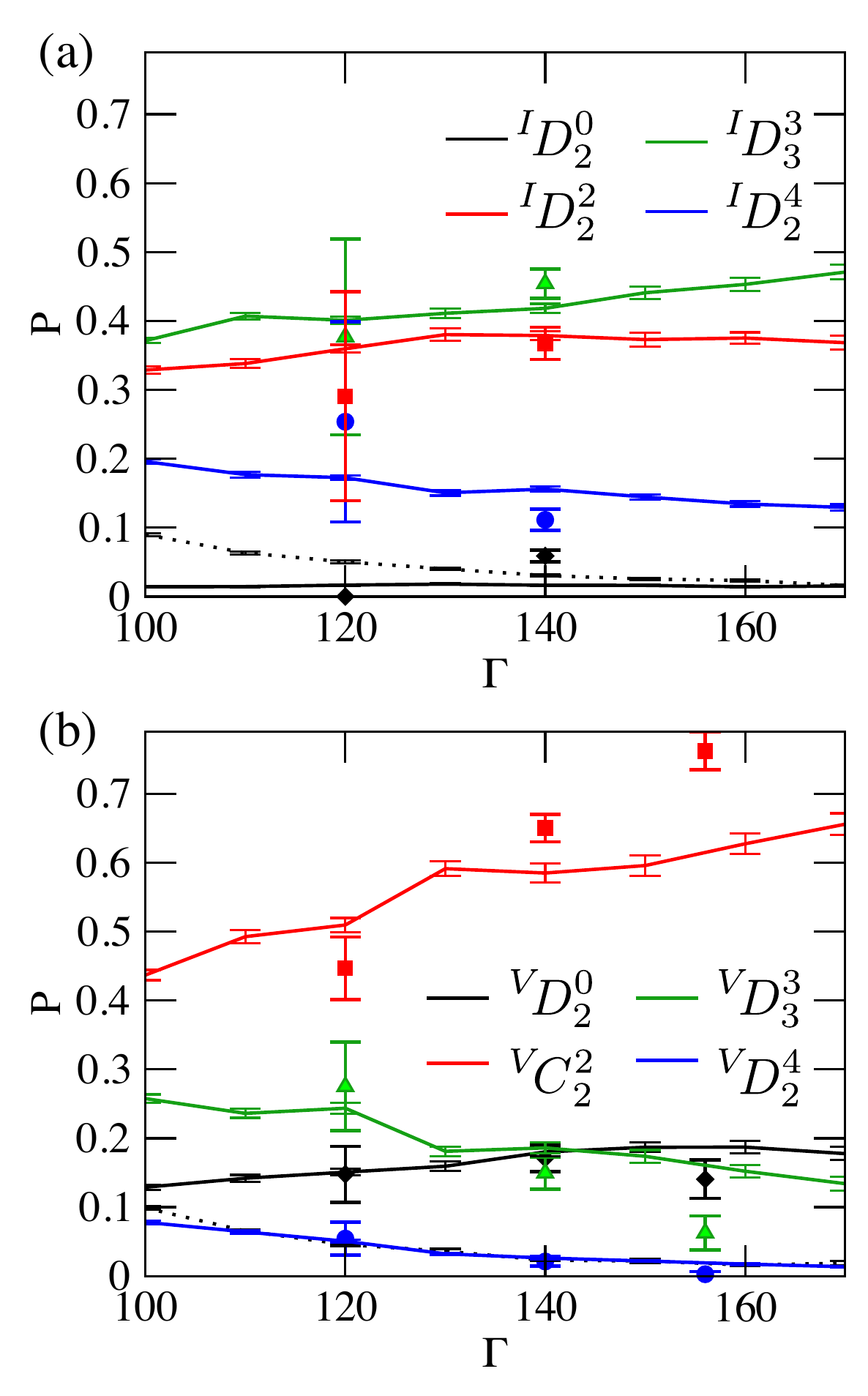}}
\caption{Equilibrium populations of (a) interstitial and (b) vacancy states identified based on Table I as a function of $\Gamma$ from experiments  (filled symbols) and simulations (solid lines). The dotted line denotes the probability to find the defect in a state not given in the table which are negligible for higher interaction strength $\Gamma$. Error bars depict the statistical error from finite numbers of measurements taking into account the correlations between the individual configurations.}
\label{fig:probabilities}
\end{figure}

Populations of interstitial and vacancy states obtained from computer simulations for $\Gamma$ ranging from $\Gamma=100$ to $170$ are shown in Fig. \ref{fig:probabilities}. The defect populations determined experimentally for interstitials and vacancies at $\Gamma=120,140$ and $156$ are in good agreement. For the vacancies the small overpopulation  for $\Gamma < 140$ of $^VC_2^2$-type with the lowest symmetry (red squares) is attributed to a tiny shear within the sample since vacancies were mainly created at the water/air interface. This interface is less stable compared to the solid substrate but vacancies can be created with less perturbations by pushing particles out of the interface using the light pressure of the laser pulse.

\section{Free Energy of Defect States}

The probability of finding the defect in state $i$ is related to the free energy of the state, which consists of an energy and an entropy contribution. For convenience we introduce the reduced energy $W(\mathbf{x}) = \frac{3^{3/4}}{2 \pi^{3/2}} \sum_{i<j} \left( a/r_{ij} \right)^3$,which is related to the potential energy $V(\mathbf{x})$ by $\beta V(\mathbf{x}) = \Gamma W(\mathbf{x})$. The probability density in configuration space for a given temperature and volume is given by
\begin{equation}
\rho(\mathbf{x}) = Z^{-1} e^{-\Gamma W(\mathbf{x})}
\end{equation}
with the partition function $Z = \int e^{-\Gamma W(\mathbf{x})} \,d\mathbf{x}$. The probability of finding the defect in state $i$ can then be written as ensemble average,
\begin{equation}
P_i = \frac{\int e^{-\Gamma W(\mathbf{x})} h_i(\mathbf{x})\, d\mathbf{x}}{\int e^{-\Gamma W(\mathbf{x})}\, d\mathbf{x}} =  \frac{Z_i}{Z} = e^{-\Gamma \Delta F_i},
\end{equation}
where $Z_i=\int e^{-\Gamma W(\mathbf{x})} h_i(\mathbf{x})\, d\mathbf{x}$ is the partition function restricted to state $i$. Thus,
$\Delta F_i = F_i-F$ is the difference between the free energy $F_i=-\Gamma^{-1}\ln Z_i$ of configuration $i$ and the free energy $F=-\Gamma^{-1} \ln Z$ of the system. According to basic statistical mechanics, the free energy $F_i$ is the sum of an energetic and an entropic contribution,
\begin{equation}
F_i = \langle W \rangle_i - \frac{S_i}{\Gamma}.
\end{equation}
Here,  $\langle W \rangle_i$ is the average energy given that the system is in state $i$, and $S_i$ is the entropy of state $i$.

\begin{figure}[htb]
\centerline{\includegraphics[width=8.5cm]{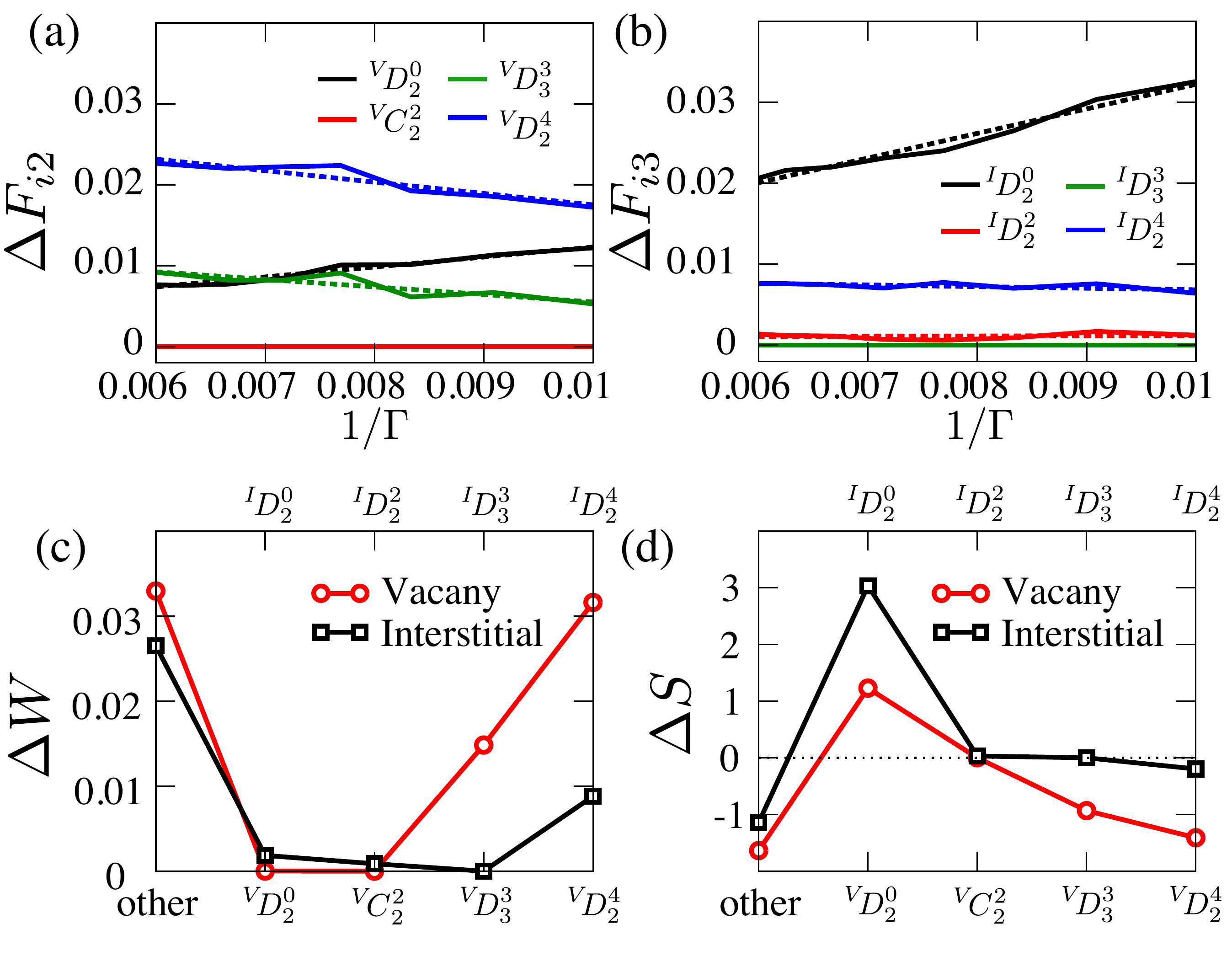}}
\caption{Free energy difference $\Delta F_{kl}$ as a function of $1/\Gamma$ for vacancies (a) and interstitials (b). The dashed lines are linear fits to the data obtained from computer simulations. According to Eq. (\ref{equ:freeenergy}), the intercepts of these functions corresponds to energy differences $\Delta W_{kl}$ shown in panel (c) for interstitials (black) and vacancies (red). The energies are given with respect to the lowest energy state which is $^VC_2^2$ and $^ID_3^3$ for vacancies and interstitials, respectively. The slopes of the lines in panels (a) and (b) yield the entropy differences $\Delta S_{kl}$ shown in panel (d) for  interstitials (black) and vacancies (red).}
\label{fig:energies}
\end{figure}

For sufficiently large values of $\Gamma$, at which configurations belonging to state $i$ can be viewed as small fluctuations about a local energy minimum, the energetic and entropic contributions to the free energy can be computed analytically. In this regime, the energy of each state $i$ is approximated as quadratic function centred at the minimum energy configuration $\mathbf{x}_i^0$,
\begin{equation}
W_i(\mathbf{x}) = W(\mathbf{x}_i^0) + \frac{1}{2} \mathbf{u}_i \mathcal{D}_i \mathbf{u}_i,
\end{equation}
where $\mathbf{u}_i = \mathbf{x} - \mathbf{x}^0$ is the displacement from the minimum and $\mathcal{D}_i$ is the matrix of second derivatives of $W$ evaluated at $\mathbf{x}_i^0$. The partition function of state $i$ is then given by
$Z_i=e^{-\Gamma W(\mathbf{x}_i^0)}\sqrt{\frac{(2\pi)^n}{\Gamma^n {\rm det} D_i}}$,
where $n$ is the number of degrees of freedom. From this expression it follows that the free energy difference $\Delta F_{kl}=F_k-F_l$ between two states $k$ and $l$, which determines the relative population $P_k/P_l$, can be expressed as
\begin{eqnarray}
\Delta F_{kl}&=& -\frac{1}{\Gamma} \ln \frac{P_k}{P_l} = \label{equ:freeenergy} \\ \nonumber
&=& W(\mathbf{x}^0_k) - W(\mathbf{x}^0_l) - \frac{1}{\Gamma} \frac{1}{2} \ln \frac{\det  \mathcal{D}_k}{\det  \mathcal{D}_l} \\ \nonumber
&=& \Delta W_{kl} - \frac{1}{\Gamma} \Delta S_{kl}.
\end{eqnarray}
Thus, the free energy difference $\Delta F_{kl}=F_k-F_l$ depends linearly on $1/\Gamma$. The intercept of this function with the $y$-axis then yields the energy difference $\Delta W_{kl}= W(\mathbf{x}^0_k) - W_l(\mathbf{x}^0_l)$ between states $k$ and $l$ and the slope equals the entropy difference $\Delta S_{kl}$. As shown in Fig. \ref{fig:energies}, this linear behavior is indeed observed in our simulations. Figure \ref{fig:energies} also shows the energy and entropy differences obtained by linear fits of Eq. (\ref{equ:freeenergy}) to the simulation results. As can be inferred from the figure the symmetric defect configurations $^VD_2^0$ and $^ID_2^0$ have a positive entropy with respect to the respective lowest energy states, leading to a negative slope of the free energy vs. $1/\Gamma$ curves shown as black lines in Fig. \ref{fig:energies} (a) and (b). Note, that while the energy difference between states is small for all $\Gamma$, the entropy vs. energy ratio may change  dramatically as a function of $\Gamma$. This positive entropy difference for the symmetric defect configurations causes an inversion of the population order for lower values of $\Gamma$.

\section{Defect Kinetics}

On a coarse level, neglecting microscopic details, the motion of a defect can be viewed as a sequence of transitions between discrete states. This type of dynamics can be mapped onto a Markov state model governed by the master equation
\begin{equation}
\frac{d P_{i}(t)}{d t} = \sum_{j\neq i} \left[ K_{ij} P_j(t) - K_{ji} P_i(t) \right].
\label{equ:generalME}
\end{equation}
Here, $P_i(t)$ is the probability of finding the defect in state $i$ at time $t$ and $K_{ij}$ is the rate constant for transitions from state $j$ to state $i$. The general solution of the master equation can be written in matrix and vector notation as
\begin{equation}
\mathbf{P}(t) = \exp[\mathbf{K} t] \mathbf{P}(0).
\label{equ:equationofmotion}
\end{equation}
Here, $\mathbf{P}(t)$ is the vector of probabilities $P_i$ at time $t$ and $\mathbf{K}$ is the matrix of transition rate constants $K_{ij}$. While for short times  $\mathbf{P}(t)$ depends on the initial conditions  $\mathbf{P}(0)$, for long times $\mathbf{P}(t)$ converges to the vector $\mathbf{P}^{\rm eq}$ of equilibrium populations independent of time and initial conditions,
\begin{equation}
\mathbf{P}^{\rm eq} = \lim\limits_{t \rightarrow \infty} \exp[\mathbf{K} t] \mathbf{P}(0),
\label{equ:pss}
\end{equation}
For the matrix of rate constants, the condition of detailed balance holds with respect to the equilibrium distribution  $\mathbf{P}^{\rm eq}$,
\begin{equation}
\frac{P^{\rm eq}_i}{P^{\rm eq}_j} =  \frac{K_{ji}}{K_{ij}}.
\label{equ:detailedbalance}
\end{equation}
In addition, the conservation of total probability requires
\begin{equation}
K_{ii} = -\sum_{j \neq i} K_{ji}.
\label{equ:conservation}
\end{equation}
The transition rate constants $K_{ij}$ can be calculated from a trajectory of states with the following procedure. To characterise the time evolution of the system, we introduce the  correlation functions
\begin{equation}
C_{ij}(t) = P(i,t|j,0) = \frac{\langle h_i[\mathbf{x}(t)] h_j\mathbf{x}(0)] \rangle}{\langle h_j[\mathbf{x}(0)] \rangle}.
\label{equ:transitions}
\end{equation}
The correlation function $C_{ij}(t)$ is the conditional probability of finding the defect in state $i$ at time $t$, given that it was in state $j$ at time $t'=0$. The equilibrium probability $P^{\rm eq}_i$ of state $i$ is the large time limit of $P(i,t|j,0)$,
\begin{equation}
P^{\rm eq}_i=\lim\limits_{t \rightarrow \infty}{P(i,t|j,0)}.
\label{equ:transitions}
\end{equation}
To obtain this equation, we have used the fact that for long times the state of the system at time $t$ is statistically independent of the state at time $t=0$, i.e., $\langle h_i(t) h_j(0) \rangle = \langle h_i(t)\rangle \langle h_j(0) \rangle$. The correlation functions $C_{ij}(t)$ can be easily determined from trajectories obtained in experiments or simulations. This correlation function is then compared to the result of the master equation given the matrix of rate constants $K_{ij}$ with
\begin{equation}
\hat{C}_{ij}(t) = \left(\exp[\mathbf{K}t]\mathbf{P}^{\textrm{init}}_j\right)_i.
\label{equ:cijmasterequation}
\end{equation}
Here, the initial vector $\mathbf{P}^{\textrm{init}}_j$ has a $1$ in component $j$ while all other components have a value of $0$. This particular choice of initial condition implies that system is initially in state $j$ with probability $1$ as required by the definition of the conditions probability $C_{ij}$.

\begin{figure}[htb]
\centerline{\includegraphics[width=8.5cm]{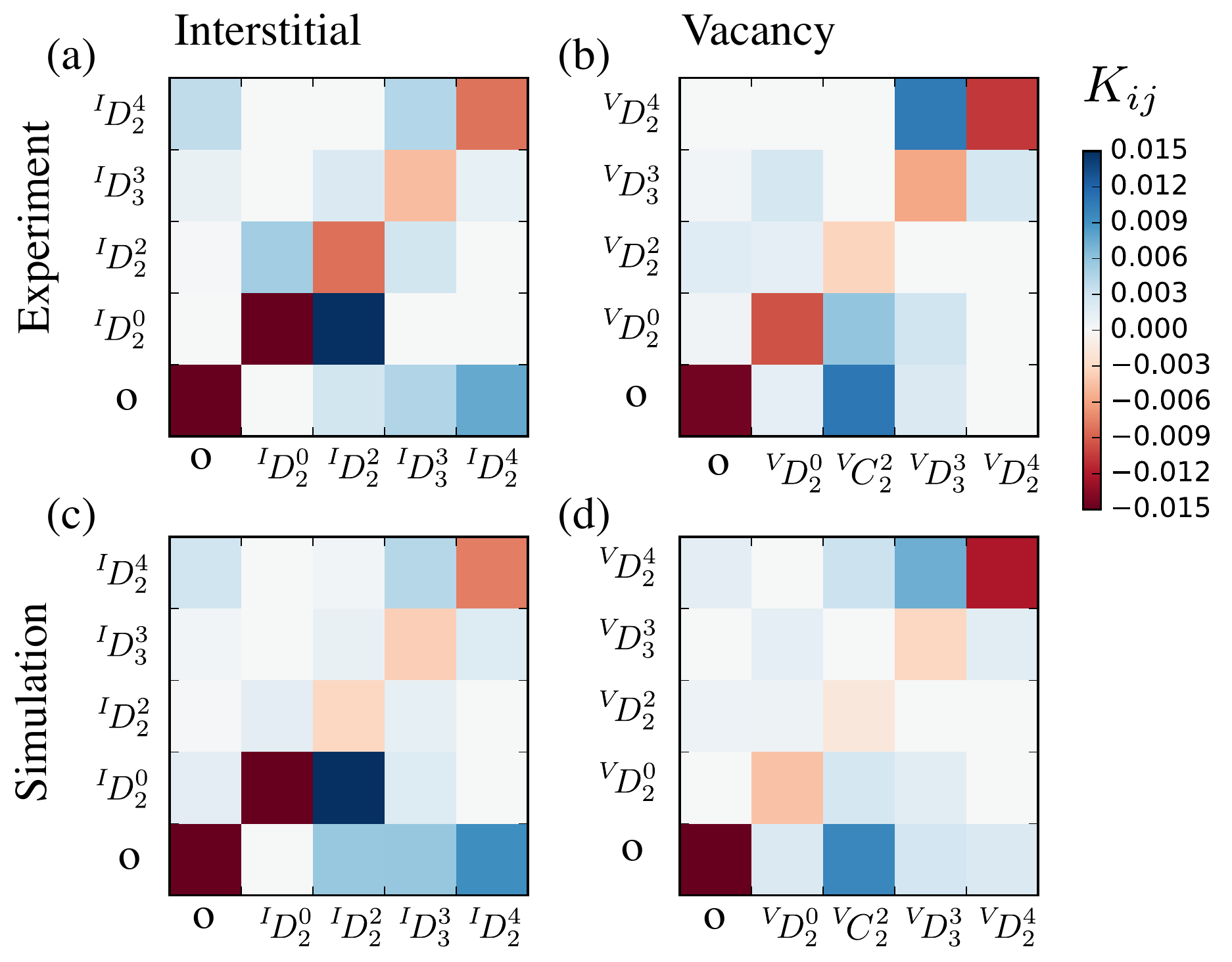}}
\caption{Color coded optimized rate matrix $\mathbf{K}_{ij}$ for an interstitial (a) and (c) and a vacancy (b) and (d) from a trajectory from experiment (top) and simulation (bottom) at $\Gamma=140$.}
\label{fig:kij}
\end{figure}

To determine the matrix of rate constants $K_{ij}$ governing the dynamics of defects, we carry out an optimization procedure that minimises the difference between the time correlation functions measured in experiments or simulations and those predicted from the solution of the master equation. The target function of this optimisation is defined as
\begin{equation}
E(\mathbf{K}) = \sum_{i,j} \sum_{l} \left[ \hat{C}_{ij}(l\Delta t) - C_{ij}(l\Delta t, \mathbf{K}) \right]^2 .
\label{eq:target}
\end{equation}
Here,  $C_{ij}(\Delta t)$ denotes the matrix of correlation functions determined from experiment or simulation. The argument $\mathbf{K}$ in the target function $E(\mathbf{K})$ and in the matrix of predicted correlation functions $ \hat{C}_{ij}(\Delta t, \mathbf{K})$ emphasise their dependence on the matrix of rate constants. The matrix of rate constants is optimized to best reproduce the observed time correlation functions. In this optimization procedure the target function of Eq. (\ref{eq:target}) is then minimized iteratively. In each step of the iteration, a matrix element $K_{ij}$ is chosen at random and changed by adding a random amount $\Delta k$. To satisfy the constraint Eq. (\ref{equ:conservation}) $\Delta k$ is also subtracted from matrix element $K_{ii}$ . The step is accepted if the target function has decreased. The iteration is stopped when the target function has not decreased for a certain number of steps. The optimisation procedure is initialized with the time derivatives of the time correlation functions $C_{ij}(t)$ evaluated at $t=0$, with $K^{TST}_{ij}= d C_{ij}(t)/ dt|_{t=0}$. These initial transition rate constants correspond to the transition state theory estimates obtained for the dividing surfaces defined implicitly by the state classification introduced earlier. In the transition state theory approximation, correlated crossings of the dividing surface are neglected. Note, that the optimization procedure described above takes these transient short-time correlations correctly into account.

\begin{table}
\label{tbl:kij}
\begin{tabular}{l}
Experiment - Interstitials [$\times 10^2$]  \\
\begin{tabular}{ccccc}
\hline
-1.5014 & 0.0000 & 0.2827 & 0.4473 & 0.7714 \\
0.0040 & -2.1958 & 2.1918 & 0.0000 & 0.0000 \\
0.0218 & 0.5257 & -0.8295 & 0.2820 & 0.0000 \\
0.1171 & 0.0000 & 0.2219 & -0.4665 & 0.1275 \\
0.3825 & 0.0000 & 0.0000 & 0.4346 & -0.8172 \\
\end{tabular}
\end{tabular}
\begin{tabular}{l}
Simulation - Interstitials  [$\times 10^2$] \\
\begin{tabular}{ccccc}
\hline
-2.0873 & 0.0000 & 0.5688 & 0.5834 & 0.9351 \\
0.1637 & -1.9832 & 1.6178 & 0.2016 & 0.0000 \\
0.0231 & 0.1631 & -0.3263 & 0.1401 & 0.0000 \\
0.0578 & 0.0000 & 0.1093 & -0.3682 & 0.2012 \\
0.2939 & 0.0000 & 0.0543 & 0.4219 & -0.7701 \\
\end{tabular}
\end{tabular}
\begin{tabular}{l}
Experiment - Vacancies [$\times 10^2$]      \\
\begin{tabular}{ccccc}
\hline
-1.4511 & 0.1469 & 1.0814 & 0.2227 & 0.0000 \\
0.0694 & -0.9595 & 0.5917 & 0.2984 & 0.0000 \\
0.1820 & 0.1521 & -0.3341 & 0.0000 & 0.0000 \\
0.0509 & 0.2643 & 0.0000 & -0.5841 & 0.2689 \\
0.0000 & 0.0000 & 0.0000 & 1.0602 & -1.0603 \\
\end{tabular}
\end{tabular}
\begin{tabular}{l}
Simulation - Vacancies [$\times 10^2$]      \\
\begin{tabular}{ccccc}
\hline
-1.7116 & 0.2209 & 0.9771 & 0.2794 & 0.2342 \\
0.0000 & -0.4375 & 0.2625 & 0.1734 & 0.0015 \\
0.0874 & 0.0812 & -0.1686 & 0.0000 & 0.0000 \\
0.0000 & 0.1483 & 0.0000 & -0.3230 & 0.1747 \\
0.1621 & 0.0000 & 0.3248 & 0.7269 & -1.2137 \\
\end{tabular}
\end{tabular}
\caption{Rate constant matrix $K_{ij}$ for $\Gamma=140$ obtained from experiment and simulation using the optimization of Eq. \ref{eq:target}.
}
\end{table}

Transition rate constants obtained with this optimization procedure for vacancies and interstitials are summarized in Fig. \ref{fig:kij}. The results from computer simulations and experiments are in very good agreement. The results show that the transition rates are not homogeneous. The most dominant transitions are $^ID_2^0 \rightarrow ^ID_2^2$ for interstitials and $^VD_2^4 \rightarrow ^VD_3^3$ for vacancies. We also identify several transitions with a vanishing rate (e.g. $^VI_2^4 \rightarrow ^ID_2^0$ or $^VD_2^4 \rightarrow ^VC_2^2$. We plan to investigate the exact microscopic mechanism for these rare  transitions in the future.\\

In summary we have shown that geometrical defects like interstitials and vacancies appear as different topological configurations mainly constructed of two, three, or four dislocations where the burger-vector cancels. We characterize the configurations by the symmetry of 2D point groups and show that the symmetries for interstitials and vacancies are not equivalent. The relative equilibrium probabilities of defects vary as a function of the temperature. In the low temperature limit, the probabilities of $^ID_2^2$ and $^ID_3^3$ symmetries are largest for interstitials, while vacancies are predominantly in the $^VC_2^2$ symmetry. This completely different temperature dependence of vacancies and interstitials is not dominated by their energy, which is almost degenerated but by the entropy. The entropic and energetic contributions can be accurately determined from a second order expansion of the energy with respect to displacements.

The kinetics of the defects is well described by a master equation in a multi-state Markov model. The states are different symmetries of the defects and the rates between different states are determined from time correlation functions,  which we measure in experiment and computer simulations.

This work presents a detailed study on the defect energetics and dynamics of point defects in two dimensional materials. We hope that this motivates experiments in other two dimensional systems. A particular future question which becomes accessible, e.g. in ultracold dipolar quantum gases \cite{MOLECULES} and graphene \cite{GRAPHENE} is the role quantum fluctuations in the defect dynamics.

{\it Acknowledgements}- Work was supported by the Austrian Science Fund (FWF): P 25454-N27 and the German Research Foundation (DFG), SFB-TR6, project C2 and SFB ViCoM (Grant No. F41).

\bibliographystyle{prsty}

\end{document}